\documentclass[showpacs,amsmath,twocolumn,showkeys,pra]{revtex4-1}
\usepackage{dcolumn}    
\usepackage{bm}         
\usepackage{ifpdf}
\usepackage{amssymb,lineno,amsfonts}
\usepackage{graphicx}   
\usepackage{bbm}
\usepackage{mathrsfs}
\usepackage{upgreek}
\usepackage{mathtools}
\usepackage{epstopdf}
\usepackage{setspace}
\usepackage{hyperref}
\usepackage[matrix,frame,arrow]{xypic}
\usepackage{rotating}
\usepackage{float}
\usepackage{natbib}
\usepackage{color} 
\definecolor{med-blue}{RGB}{25,25,112} 
\hypersetup{colorlinks, linkcolor={red},citecolor={blue}, urlcolor={blue}}

\begin{document}
\vspace*{1cm}
\title{Ancilla Assisted Quantum State Tomography in Many-Qubit Registers}
\author{Abhishek Shukla$^1$, K. Rama Koteswara Rao$^2$, and T. S. Mahesh$^1$}
\email{mahesh.ts@iiserpune.ac.in}
\affiliation{$ ^1 $Department of Physics and NMR Research Center,\\
Indian Institute of Science Education and Research, Pune 411008, India \\
$ ^2 $Department of Physics and NMR Research Centre,
Indian Institute of Science, Bangalore, India}

\begin{abstract}
{
The standard method of Quantum State Tomography (QST) relies on the
measurement of a set of noncommuting observables, realized in 
a series of independent experiments. Ancilla Assisted QST (AAQST) proposed 
by Nieuwenhuizen and co-workers 
(Phys. Rev. Lett., {\bf 92}, 120402 (2004))
greatly reduces the number of independent
measurements by exploiting an ancilla register in a known initial state.
In suitable conditions AAQST allows mapping out 
density matrix of an input register in a single experiment.
Here we describe methods for explicit construction of 
AAQST experiments in multi-qubit registers.
We also report nuclear magnetic resonance studies on  
AAQST of (i) a two-qubit input register using a one-qubit ancilla 
in an isotropic liquid-state system and (ii) a three-qubit input register 
using a two-qubit ancilla register
in a partially oriented system.  The experimental results confirm the
effectiveness of AAQST in such many-qubit registers.
}
\end{abstract}

\maketitle

\section{Introduction}
Quantum computers have the potential to carry-out certain computational tasks 
with an efficiency that is beyond the reach of their classical
counterparts \cite{chuangbook}.  In practice however, harnessing the computational 
power of a quantum system has been an enormously challenging task
\cite{exptqipbook}.
The difficulties include imperfect control on the quantum dynamics
and omnipresent interactions between the quantum system and its
environment leading to an irreversible loss of quantum coherence.
In order to optimize the control fields and to understand the effects
of environmental noise, it is often necessary to completely characterize the quantum state.
In experimental quantum information studies,
Quantum State Tomography (QST) is an important tool that is routinely used
to characterize an instantaneous quantum state \cite{chuangbook}.  

QST on an initial state is usually carried out to confirm the efficiency of
initialization process.  Though QST of the final state is usually not  
part of a quantum algorithm, it allows one
to measure the fidelity of the output state.  
QSTs in intermediate stages often help experimentalists
to tune-up the control fields better.  

QST can be performed by a series of measurements of
noncommuting observables which together enables one to reconstruct 
the complete complex density matrix.  In the standard method,
the required number of independent experiments grows
exponentially with the number of input qubits \cite{ChuangPRSL98,ChuangPRA99}.
Anil Kumar and co-workers have illustrated QST using a single
two-dimensional NMR spectrum \cite{Aniltomo}.  They showed that a two-dimensional
NMR experiment consisting of a series of identical measurements with systematic
increments in evolution time, can be used to quantitatively estimate all
the elements of the density matrix.  
Later Nieuwenhuizen and co-workers have shown that 
it is possible to reduce the number of independent experiments
in the presence of an ancilla register initialized to a known
state \cite{Nieuwenhuizen}.
They pointed out that in suitable situations, it is possible to 
carry-out QST with a single measurement of a set of factorized
observables.  We refer to this method as Ancilla Assisted QST (AAQST).
This method was experimentally illustrated by Suter
and co-workers using a single input qubit and a single ancilla 
qubit \cite{sutertomo}.  Recently Peng and coworkers have studied the effectiveness
of the method for qutrit-like systems using numerical simulations 
\cite{pengtomo}.  

Single shot mapping of density matrix by AAQST method not only
reduces the experimental time, but also alleviates the need to 
prepare the target state several times.  Often slow variations
in system Hamiltonian may result in systematic errors in 
repeating the state preparation.  Further, environmental noises
lead to random errors in multiple preparations.  
These errors play important roles in the quality of the 
reconstruction of the target state. Therefore AAQST has the 
potential to provide a more reliable way of tomography.  

In this article we first revisit the theory of QST and AAQST and provide
methods for explicit construction of the constraint matrices,
which will allow extending the tomography procedure for large registers.
An important feature of the method described here is that
it requires only global rotations and short evolutions under the collective 
internal Hamiltonian.  
We also describe nuclear magnetic resonance (NMR) demonstrations 
of AAQST on two different types of systems:
(i) a two-qubit input register using a one-qubit ancilla in an isotropic liquid-state system 
and (ii) a three-qubit input register using a two-qubit ancilla register
in a partially oriented system. 

In the following section we briefly describe the theory of QST and AAQST.
In section III we describe experimental demonstrations  and finally we conclude in section IV.

\section{Theory}
\subsection{Quantum State Tomography}
We consider an $n$-qubit register 
formed by a system of $n$ mutually interacting spin-1/2 nuclei
with distinct resonance frequencies $\omega_i$ and mutual
interaction frequencies $2\pi J_{ij}$.
The Hamiltonian under weak-interaction limit ($2 \pi J_{ij}\ll \vert \omega_i-\omega_j \vert$) 
consists of the Zeeman part and spin-spin interaction part, i.e.,
\begin{eqnarray}
{\cal H} = -\sum\limits_{i=1}^{n}\omega_i \sigma_z^i /2 + 
\sum\limits_{i=1}^{n}\sum\limits_{j=i+1}^{n} 2\pi J_{ij} \sigma_z^i \sigma_z^j /4
\label{ham}
\end{eqnarray}
respectively, where $\sigma_{z}^i$ and $\sigma_{z}^j$ are the $z$-components of Pauli operators
of $i$th and $j$th qubits \cite{cavanagh}.  
The set of $N=2^n$ eigenvectors $\{ \vert m_1 m_2 \cdots m_n \rangle \}$ 
of the Zeeman Hamiltonian form a complete orthonormal computational basis.  
We can order the eigenvectors based on the decimal value $m$
of the binary string $(m_1 \cdots m_n)$, i.e., $m = m_1 2^{n-1}+\cdots+m_n 2^0$.

The general density matrix can be decomposed as
$\mathbbm{1}/N+\epsilon \rho$ where the identity part is known as the background,
the trace-less part $\rho$ is known as the \textit{deviation density matrix},
and the dimensionless constant $\epsilon$ is the purity factor 
\cite{corypps}.
In this context, QST refers to complete characterization of the deviation density
matrix, which can be expanded in terms of $N^2-1$ real unknowns:
\begin{eqnarray}
\rho &=& 
\sum\limits_{m=0}^{N-2} \rho_{mm}(\vert m \rangle \langle m \vert -\vert N-1 \rangle \langle N-1 \vert)  \nonumber \\
&&+ \sum_{m=0}^{N-2}\sum_{m'=m+1}^{N-1} \{
R_{mm'}(\vert m \rangle \langle m' \vert+\vert m' \rangle \langle m \vert)\nonumber \\
&&~~~~~~~~~~~~~~~~+  iS_{mm'}(\vert m \rangle \langle m' \vert-\vert m' \rangle \langle m \vert)
\}.
\label{dmm}
\end{eqnarray}
Here first part consists of $N-1$ diagonal unknowns $\rho_{mm}$ with
the last diagonal element $\rho_{N-1,N-1}$ being constrained by the trace-less condition.  
$R$ and $S$
each consisting of $(N^2-N)/2$ unknowns correspond to real and imaginary 
parts of the off-diagonal elements respectively.
Thus a total of $N^2-1$ real unknowns needs to be determined.

Usually an experimental technique allows a particular set of observables 
to be measured directly.
To explain the NMR case, we introduce $n$-bit binary strings,
$j_\nu = \nu_1 \nu_2 \cdots \nu_{j-1} 0 \nu_{j} \cdots \nu_{n-1}$ and
$j'_{\nu} = \nu_1 \nu_2 \cdots \nu_{j-1} 1 \nu_{j} \cdots \nu_{n-1}$
differed only by the flip of the $j$th bit.
Here $\nu = \nu_1 2^{n-2} + \nu_2 2^{n-3} + \cdots + \nu_{n-1} 2^0$ is the 
value of the $n-1$ bit binary string $(\nu_1,\nu_2,\cdots,\nu_{n-1})$
and $\nu$ can take a value between $0$ and $\gamma = N/2-1$.
The real and imaginary parts of an NMR signal recorded in a 
quadrature mode corresponds to the expectation values of 
transverse magnetization observables $\sum\limits_{j=1}^{n}\sigma_{jx}$ and 
$\sum\limits_{j=1}^{n}\sigma_{jy}$ respectively \cite{cavanagh}.

The background part of the density matrix neither evolves under
unitaries nor gives raise to any signal, and therefore we ignore it.
Under suitable conditions (when all the transitions are resolved), 
a single spectrum directly yields $nN$
matrix elements $\{R_{j_\nu,j_\nu'},S_{j_\nu,j_\nu'}\}$
as complex intensities of spectral lines.  These matrix elements
are often referred to as single quantum elements since they
connect eigenvectors related by the flip of a single qubit.  
We refer the single-quantum terms $R_{j_\nu,j_\nu'}$
and $S_{j_\nu,j_\nu'}$ respectively
as the real and imaginary parts of $\nu$th spectral line of $j$th qubit.
Thus a single spectrum of an $n$-qubit system in an
arbitrary density matrix can yield $n N$ real unknowns. 

In order to quantify the remaining elements,
one relies on multiple experiments all starting from the same initial state
$\rho$.  The $k$th experiment consists of applying a unitary $U_k$
to the state $\rho$, leading to $\rho^{(k)} = U_k \rho U_k^\dagger$, and measuring 
the single-quantum spectrum $\{R_{j_\nu,j_\nu'}^{(k)},S_{j_\nu,j_\nu'}^{(k)}\}$. 
From eqn. (\ref{dmm}) we obtain
\begin{eqnarray}
&&R^{(k)}_{j_\nu,j_\nu'} 
 = \sum\limits_{m} a_{j\nu}^{(k)}(m) \rho_{mm} + 
\nonumber \\
&& ~~~~~~~ \sum\limits_{m,m'>m}c_{j\nu}^{(k)}(m,m') R_{mm'} + e_{j\nu}^{(k)}(m,m') S_{mm'}, 
\nonumber \\
&&S^{(k)}_{j_\nu,j_\nu'} 
 = \sum\limits_{m} b_{j\nu}^{(k)}(m) \rho_{mm} + 
 \nonumber \\
&& ~~~~~~~ \sum\limits_{m,m'>m} d_{j\nu}^{(k)}(m,m') R_{mm'} + f_{j\nu}^{(k)}(m,m') S_{mm'}, 
\label{leq}
\end{eqnarray}
in terms of the unknowns $\rho_{mm'}$ and the known real constants $\{a, \cdots, f\}$:
\begin{eqnarray}
a_{j\nu}^{(k)}(m,m) + ib_{j\nu}^{(k)}(m,m) &=& 
\langle j_\nu \vert U_k \vert m \rangle \langle m \vert U_k^\dagger \vert j'_\nu\rangle-
\nonumber \\
&&\langle j_\nu \vert U_k \vert N-1 \rangle \langle N-1 \vert U_k^\dagger \vert j'_\nu\rangle, 
\nonumber \\
c_{j\nu}^{(k)}(m,m') +i d_{j\nu}^{(k)}(m,m')&=& 
\langle j_\nu \vert U_k \vert m \rangle \langle m' \vert U_k^\dagger \vert j'_\nu\rangle + 
\nonumber \\
&&\langle j_\nu \vert U_k \vert m' \rangle \langle m \vert U_k^\dagger \vert j'_\nu\rangle, 
\nonumber \\
e_{j\nu}^{(k)}(m,m')+if_{j\nu}^{(k)}(m,m') &=& 
i\langle j_\nu \vert U_k \vert m \rangle \langle m' \vert U_k^\dagger \vert j'_\nu\rangle - 
\nonumber \\
&&i\langle j_\nu \vert U_k \vert m' \rangle \langle m \vert U_k^\dagger \vert j'_\nu\rangle
\end{eqnarray}
\cite{maheshtomo}.
After $K$ experiments, we can setup the matrix equation
\begin{eqnarray}
M
\left[
\begin{array}{c}
\rho_{0,0} \\
\cdots \\
\rho_{N-2,N-2} \\
-------- \\
R_{0,1} \\
\cdots \\
R_{0,N-1} \\
\cdots \\
R_{m,m'>m} \\
\cdots \\
R_{N-2,N-1} \\
-------- \\
S_{0,1} \\
\cdots \\
S_{0,N-1} \\
\cdots \\
S_{m,m'>m} \\
\cdots \\
S_{N-2,N-1} \\
\end{array}
\right]
 = 
 \left[
 \begin{array}{c}
R^{(1)}_{1_0,1_0'}  \\
 \cdots  \\
R^{(1)}_{1_\gamma,1_\gamma'}  \\
R^{(1)}_{2_0,2_0'}  \\
 \cdots  \\
 \cdots  \\
R^{(K)}_{n_\gamma,n_\gamma'} \\
-------\\
S^{(1)}_{1_0,1_0'}  \\
 \cdots  \\
S^{(1)}_{1_\gamma,1_\gamma'}  \\
S^{(1)}_{2_0,2_0'}  \\
 \cdots  \\
 \cdots  \\
S^{(K)}_{n_\gamma,n_\gamma'} \\
 \end{array}
 \right].
\label{meq}
\end{eqnarray}
Here the left column vector is formed by the $N^2-1$ unknowns of $\rho$:
diagonal elements in the top, real off-diagonals in the middle,
and imaginary off-diagonals in the bottom.  
The right column vector is formed by $KnN$ numbers - 
the real and imaginary parts of the experimentally obtained 
spectral intensities ordered according to the value of the 
binary string $\nu$, the qubit number $j$, and the experiment number $k$.
The $KnN\times(N^2-1)$ dimensional constraint matrix is of the form
\begin{eqnarray}
&&M = \nonumber \\
&&\left[
\begin{array}{c c|c c|c c}
a_{1,0}^{(1)}(0,0) & \cdots & c_{1,0}^{(1)}(m,m') & \cdots  & e_{1,0}^{(1)}(m,m') & \cdots \\
\cdots &\cdots &\cdots &\cdots &\cdots &\cdots   \\
a_{1,\gamma}^{(1)}(0,0) & \cdots & c_{1,\gamma}^{(1)}(m,m') & \cdots  & e_{1,\gamma}^{(1)}(m,m') & \cdots \\
\cdots &\cdots &\cdots &\cdots &\cdots &\cdots   \\
a_{n,0}^{(1)}(0,0) & \cdots & c_{n,0}^{(1)}(m,m') & \cdots  & e_{n,0}^{(1)}(m,m') & \cdots \\
\cdots &\cdots &\cdots &\cdots &\cdots &\cdots   \\
\cdots &\cdots &\cdots &\cdots &\cdots &\cdots   \\
a_{n\gamma}^{(K)}(0,0) & \cdots & c_{n\gamma}^{(K)}(m,m') & \cdots  & e_{n\gamma}^{(K)}(m,m') & \cdots \\
\hline
b_{1,0}^{(1)}(0,0) & \cdots & d_{1,0}^{(1)}(m,m') & \cdots  & f_{1,0}^{(1)}(m,m') & \cdots \\
\cdots &\cdots &\cdots &\cdots &\cdots &\cdots   \\
b_{1,\gamma}^{(1)}(0,0) & \cdots & d_{1,\gamma}^{(1)}(m,m') & \cdots  & f_{1,\gamma}^{(1)}(m,m') & \cdots \\
\cdots &\cdots &\cdots &\cdots &\cdots &\cdots   \\
b_{n,0}^{(1)}(0,0) & \cdots & d_{n,0}^{(1)}(m,m') & \cdots  & f_{n,0}^{(1)}(m,m') & \cdots \\
\cdots &\cdots &\cdots &\cdots &\cdots &\cdots   \\
\cdots &\cdots &\cdots &\cdots &\cdots &\cdots   \\
b_{n\gamma}^{(K)}(0,0) & \cdots & d_{n\gamma}^{(K)}(m,m') & \cdots  & f_{n\gamma}^{(K)}(m,m') & \cdots \\
\end{array}
\right]\nonumber .\\
\label{mmat}
\end{eqnarray}
Note that each column of the constraint matrix corresponds to
contribution of a particular unknown element of $\rho$  to
the various spectral intensities.

\begin{center}
\begin{figure}[t]
\includegraphics[width=7cm]{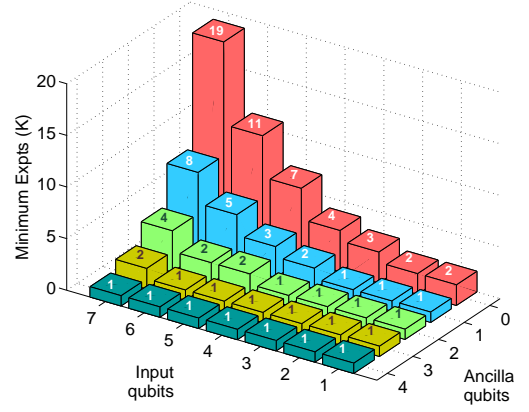} 
\caption{Minimum number of independent experiments required for QST
(with zero ancilla) and AAQST.  
}
\label{exptscaling} 
\end{figure}
\end{center}

By choosing the unitaries $\{U_k\}$ such that 
$\mathrm{rank} (M) \ge N^2-1$ (the number of unknowns),
eqn.  (\ref{meq}) can be solved either by singular value
decomposition or by Gaussian elimination method
\cite{maheshtomo}. 
Fig. \ref{exptscaling} illustrates the minimum number ($K$) 
of experiments required for QST. As anticipated, $K$ increases 
rapidly as $O(N/n)$ with the number of input qubits.
In the following we describe how it is possible to speed-up QST,
in the presence of an ancilla register, with fewer experiments.

\subsection{Ancilla Assisted QST (AAQST)}
Suppose the input register of $n$-qubits is associated with an 
ancilla register consisting of $\hat{n}$ qubits. 
The dimension of the combined system of $\tilde{n} = n+\hat{n}$ qubits 
is $\tilde{N} = N\hat{N}$, where $\hat{N} = 2^{\hat{n}}$.
For simplicity we assume that each qubit interacts sufficiently with all other qubits so as to obtain a completely resolved spectrum yielding $\tilde{n}\tilde{N}$ real parameters.
Following method is applicable even if there are spectral overlaps, 
albeit with lower efficiency (i.e., with higher number $(K)$ of minimum
experiments).  Further for simplicity, 
we assume that the ancilla register
begins with the maximally mixed initial state, with no contribution
to the spectral lines from it.  
Otherwise, we need to add the contribution of the ancilla to
the final spectrum and the eqn. (\ref{meq}) will become inhomogeneous.
As explained later in the experimental
section, initialization of maximally mixed state can be achieved with high precision.
Thus the deviation density matrix of the combined system is $\tilde{\rho} = \rho \otimes \mathbbm{1}/\hat{N}$.
Now applying only local unitaries neither leads to ancilla coherences
nor transfers any of the input coherences to ancilla.  
Therefore we consider applying a non-local unitary 
exploiting the input-ancilla interaction,
\begin{eqnarray}
\tilde{U}_k = V \sum\limits_{a=0}^{\hat{N}-1} U_{ka} \otimes \vert a \rangle \langle a \vert,
\end{eqnarray}
where $U_{ka}$ is the $k$th unitary on the input register dependent 
on the ancilla state $\vert a \rangle$ and $V$ is the local 
unitary on the ancilla.
The combined state evolves to
\begin{eqnarray}
&& \tilde{\rho}^{(k)} = \tilde{U}_k \tilde{\rho} \tilde{U}_k^\dagger \nonumber \\
&& = \frac{1}{\hat{N}}\sum\limits_{m,m',a} \rho_{mm'} U_{ka} \vert m \rangle \langle m' \vert U_{ka}^\dagger
\otimes V \vert a \rangle \langle a \vert V^\dagger.
\end{eqnarray}
We now record the spectrum of the combined system corresponding
to the observable $\sum\limits_{j=1}^{\tilde{n}} \sigma_{jx}+i\sigma_{jy}$.
Each spectral line can again be expressed in terms of the unknown
elements of the ancilla matrix in the form given in eqn. (\ref{leq}).
The spectrum of the combined system yields $\tilde{n}\tilde{N}$ linear
equations.  The minimum number of independent experiments
needed is now $O(N^2/(\tilde{n}\tilde{N}))$.
Since we can choose $\tilde{N} \gg N$, AAQST needs 
fewer than O($N/n$) experiments required in the standard QST.
In particular, when $\tilde{n}\tilde{N} \ge N^2$, a single
optimized unitary suffices for QST.
Fig. \ref{exptscaling} illustrates the minimum number ($K$) 
of experiments required for various sizes of input and ancilla 
registers. As illustrated, QST can be achieved with only one
experiment, if an ancilla of sufficient size is provided along with.

\subsection{Building the constraint matrix}
The major numerical procedure in AAQST is obtaining the constraint matrix $M$.
For calculating the constraint coefficients $c_{rj}^{(k)}$,
one may utilize an elaborate decomposition of $U_k$
using numerical or analytical methods.
Alternatively, as described below, we can use a simple algorithmic approach to 
construct the constraint matrix.

First imagine a diagonal state $\rho$ for the ancilla register
(eqn. (\ref{dmm})) with $\rho_{00}=1$ and $\rho_{mm}=0$ for all
other $1 \le m \le N-2$, $R_{mm'}=S_{mm'}=0$.  
Applying the unitary $U_k$ on the composite deviation density matrix 
$\tilde{\rho} = \rho \otimes \mathbbm{1}/\hat{N}$, we obtain all the
spectral intensities (using eqn. (\ref{leq}))
\begin{eqnarray}
a^{k}_{j\nu}(0,0) = R^{(k)}_{j\nu,j\nu'}, ~
b^{k}_{j\nu}(0,0) = S^{(k)}_{j\nu,j\nu'}.
\end{eqnarray}
Thus the spectral lines indicate the contributions only from $\rho_{00}$
(and $\rho_{N-1,N-1}$).
Repeating
the process with all the unitaries $\{U_k\}$ yields
the first column in $M$ matrix (eqn. (\ref{mmat})) corresponding to the unknown $\rho_{00}$.
Same procedure can be used for all the diagonal elements $\rho_{mm}$ with
$0 \le m \le N-2$.  

To determine $M$ matrix column corresponding to a
real off-diagonal unknown $R_{mm'}$, 
we start with an input-register density matrix 
$R_{mm'} = 1$ and all other elements  
set to zero.  Again by applying the unitary 
$U_k$ on the composite density matrix, and
using eqn. (\ref{leq}) we obtain
\begin{eqnarray}
c^{k}_{j\nu}(m,m') = R^{(k)}_{j\nu,j\nu'}, ~
d^{k}_{j\nu}(m,m') = S^{(k)}_{j\nu,j\nu'}.
\end{eqnarray}
Repeating the process with all unitaries $\{U_k\}$ 
determines the column of $M$ corresponding to the unknown $R_{mm'}$. 

To determine $M$ matrix column corresponding to 
an imaginary off-diagonal unknown $S_{mm'}$, 
we set $S_{mm'} = 1$ and all other elements to zero, 
and apply $U_k$ on the composite state to obtain
\begin{eqnarray}
e^{k}_{j\nu}(m,m') = R^{(k)}_{j\nu,j\nu'}, ~
f^{k}_{j\nu}(m,m') = S^{(k)}_{j\nu,j\nu'}.
\end{eqnarray}
Proceeding this way, by selectively setting the unknowns
one by one, the complete constraint matrix can be built easily.

\subsection{Optimization of Unitaries}
Solving the matrix equation (\ref{meq}) requires that
$\mathrm{rank}(M) \ge N^2-1$, the number of unknowns.
But having the correct rank is not sufficient.
The matrix $M$ must be well conditioned in order to ensure that small
errors in the observed intensities $\{R^{(k)}_{j\nu,j\nu'},S^{(k)}_{j\nu,j\nu'}\}$ 
do not contribute to large errors in the values of the elements $\rho_{mm'}$.  
The quality of
the constraint matrix can be measured by a scalar quantity called condition number
$C(M)$  defined as the ratio of the largest singular
value of $M$ to the smallest \cite{bau}.  Smaller the value of $C(M)$, better conditioned
is the constraint matrix $M$ for solving the unknowns.  Thus the condition 
number provides a convenient scalar quantity to optimize the set $\{U_k\}$
of unitaries to be selected for QST.  As explained in the experimental
section, we used a simple unitary model $U_1(\tau_1,\tau_2)$ as an initial
guess and used genetic algorithm to minimize the condition number and 
optimize the parameters $(\tau_1,\tau_2)$.

The necessary number ($K$) of independent experiments
is decided by the rank of the constraint matrix and the desired precision.
The rank condition requires that $KnN \ge N^2-1$. Introducing
additional experiments renders the problem over-determined, thus 
reducing the condition number and increasing the precision.  
In the following section
we describe the experimental results of AAQST for registers with (i) $n=2,\hat{n}=1,\tilde{n}=3$ and
(ii) $n=3,\hat{n}=2,\tilde{n}=5$ respectively.

\section{Experiments}
We report experimental demonstrations of AAQST on two 
spin-systems of different sizes and environments.  
In each case, we have chosen two density matrices for tomography.
All the experiments described below are carried out on a Bruker 500 MHz
spectrometer at an ambient temperature of 300 K 
using high-resolution nuclear magnetic resonance techniques.  

\begin{center}
\begin{figure}[t]
\includegraphics[width=6cm]{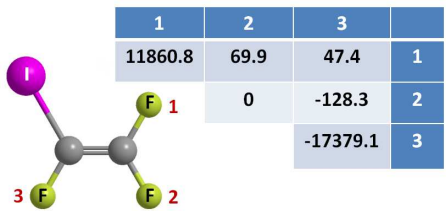} 
\caption{Molecular structure of iodotrifluoroethylene, and
the table of Hamiltonian parameters in Hz:
chemical shifts (diagonal elements) and J-coupling constants
(off-diagonal elements).
}
\label{fffmol} 
\end{figure}
\end{center}

\subsection{Two-qubit input, One-qubit ancilla}
\begin{center}
\begin{figure}[t]
\hspace*{-0.5cm}
\includegraphics[trim=0.8cm 0.8cm 0.8cm 0.2cm, clip=true,width=10cm]{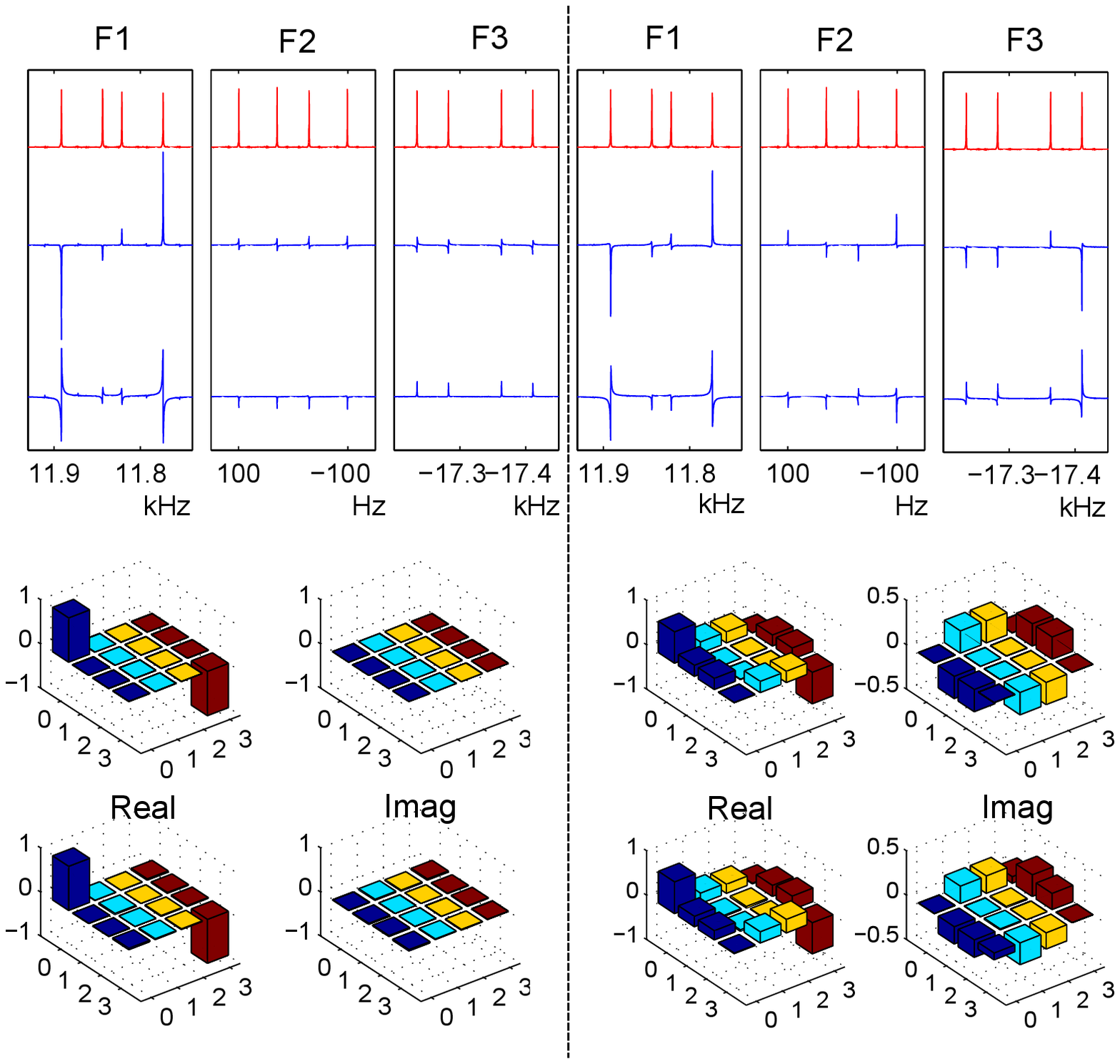} 
\caption{
AAQST results for
thermal equilibrium state $\rho_1$ (left column), and that of state $\rho_2$
(right column), described in the text.  
The reference spectra is in the top trace.  
 The spectra
corresponding to the real part ($R_{j\nu,j\nu'}^{(1)}$, middle trace)
and the imaginary part ($S_{j\nu,j\nu'}^{(1)}$, bottom trace) 
of the $^{19}$F signal are obtained in a single shot AAQST experiment.
The bar plots correspond to theoretically expected
states (top row) and those obtained from AAQST 
experiments (bottom row).
Fidelities of the states are 0.997 and 0.99 respectively
for the two density matrices.
}
\label{fffres} 
\end{figure}
\end{center}

Here we use three spin-1/2 $^{19}$F nuclei of 
iodotrifluoroethylene (C$_2$F$_3$I) dissolved in acetone-D$_6$
as a 3-qubit system.
The molecular structure and the Hamiltonian parameters are shown
in Fig. \ref{fffmol}.  
As can be seen in Fig.\ref{fffres}, all the 12 transitions 
of this system are clearly resolved.

We have chosen $F_1$ as the ancilla
qubit and $F_2$ and $F_3$ as the input qubits.
QST was performed for two different density matrices
(i) thermal equilibrium state, i.e., 
$\rho_1 = \frac{1}{2} \left( \sigma_z^2+\sigma_z^3 \right)$,
and 
(ii) state after a $(\pi/4)_{\pi/4}$ pulse applied
to the thermal equilibrium state, i.e.,
$
\rho_2 =
\frac{1}{2}\left(\sigma_x^2+\sigma_x^3\right)
-\frac{1}{2}\left(\sigma_y^2+\sigma_y^3\right)
+\frac{1}{\sqrt{2}}\left(\sigma_z^2+\sigma_z^3\right)
$.
In both the cases, the first qubit was initialized
into a maximally mixed state by applying a selective $(\pi/2)_y$
pulse on $F_1$ and followed by a strong pulsed-field-gradient (PFG)
in the $z$-direction.
The selective pulse was realized by GRAPE technique \cite{Khaneja}.

AAQST of each of the above density matrices required just one 
unitary evolution followed by the measurement of complex NMR signal. 
We modelled the AAQST unitary as follows: 
$U_1 = \left(\frac{\pi}{2}\right)_y U_\mathrm{int}(\tau_2) 
\left(\frac{\pi}{2}\right)_x U_\mathrm{int}(\tau_1)$,
where $U_\mathrm{int}(\tau) = \exp\left(-i{\cal H}\tau\right)$ is
the unitary operator for evolution under the internal Hamiltonian
${\cal H}$ (see eqn. (\ref{ham})) for a time $\tau$, and $\left(\frac{\pi}{2}\right)$ rotations
are realized by non selective radio frequency pulses applied to all the spins
along the directions indicated by the subscripts.  
The constraint matrix $M$ had 15 columns corresponding
to the unknowns and 24 rows corresponding to the real and imaginary
parts of the 12 spectral lines.
Only the durations
$\left\{\tau_1,\tau_2 \right\}$ needed to be optimized to minimize the condition
number $C(M)$.  We used a genetic algorithm for the optimization and obtained
$C(M) = 17.3$ for $\tau_1 =  6.7783$ ms and $\tau_2 = 8.0182$ ms.
The real and imaginary parts of the single shot experimental AAQST spectrum,
along with the reference spectrum,
are shown in the top part of Fig. \ref{fffres}.  The intensities
$\{R_{j\nu,j\nu'}^{(1)},S_{j\nu,j\nu'}^{(1)}\}$
were obtained by simple curve-fit routines, and the matrix eqn. (\ref{meq})
was solved to obtain all the unknowns.  The reconstructed density matrices
along with the theoretically expected ones are shown below the spectra in 
Fig. \ref{fffres}. 
The fidelities of experimental states with
the theoretically expected states ($\rho_1$ and $\rho_2$)
are respectively 0.998 and 0.990.  The high fidelities indicated successful
AAQST of the prepared states.

\begin{center}
\begin{figure}[b]
\includegraphics[width=6.5cm]{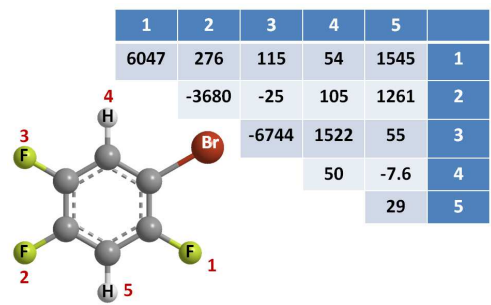} 
\caption{Molecular structure of 1-bromo-2,4,5-trifluorobenzene, and
the table of Hamiltonian parameters in Hz:
chemical shifts (diagonal elements) and effective
coupling constants (J+2D)(off-diagonal elements).
}
\label{btfbzmol} 
\end{figure}
\end{center}

\subsection{Three-qubit input, Two-qubit ancilla}
We use three $^{19}$F nuclei and two $^1$H nuclei
of 1-bromo-2,4,5-trifluorobenzene partially oriented
in a liquid crystal namely, N-(4-methoxybenzaldehyde)-4-
butylaniline (MBBA). Due to the partial orientational
order, the direct spin-spin interaction (dipolar interaction) 
does not get fully averaged out, but gets scaled down
by the order parameter \cite{dongbook}.  The chemical shifts and the 
strengths of the effective couplings are shown in
Fig. \ref{btfbzmol}. As is evident, the partially oriented
system can display stronger and longer-range coupling
network leading to a larger register.  Here we choose 
the three $^{19}$F nuclei forming the input register
and two $^1$H nuclei forming the ancilla register.
The Hamiltonian for the heteronuclear dipolar interaction 
(between $^1$H and $^{19}$F)
has an identical form as that of J-interaction \cite{dongbook}.
The homonuclear dipolar couplings 
(among $^{19}$F, as well as among $^{1}$H nuclei) were small
compared to their chemical shift differences enabling us to approximate
the Hamiltonian in the form of eqn. (\ref{ham}).

\begin{center}
\begin{figure}[t]
\includegraphics[trim=1cm 2cm 2.5cm 0cm, clip=true,width=8.5cm]{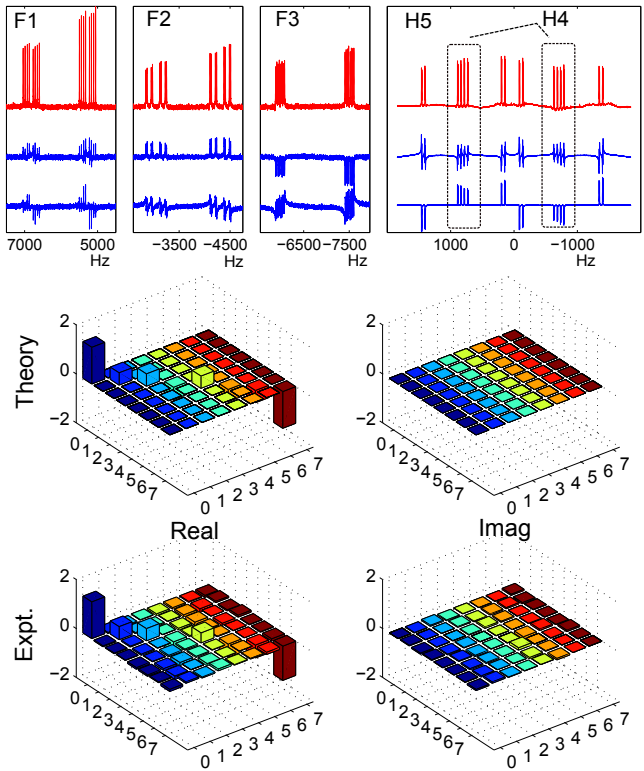} 
\caption{
AAQST results for
thermal equilibrium state, i.e., $(\sigma_z^1+\sigma_z^2+\sigma_z^3)/2$.
The reference spectrum is in the top trace.  The spectra
corresponding to the real part ($R_{j\nu,j\nu'}^{(1)}$, middle trace)
and the imaginary part ($R_{j\nu,j\nu'}^{(1)}$, bottom trace) 
of the $^{19}$F signal are obtained in a single shot AAQST experiment.
The bar plots correspond to theoretically expected
states (top row) and those obtained from AAQST 
experiments (bottom row). Fidelity of the AAQST state is 0.98.
}
\label{btfbzres1} 
\end{figure}
\end{center}
\begin{center}
\begin{figure}[t]
\includegraphics[trim=0.7cm 1.8cm 1.2cm 0cm, clip=true,width=8.5cm]{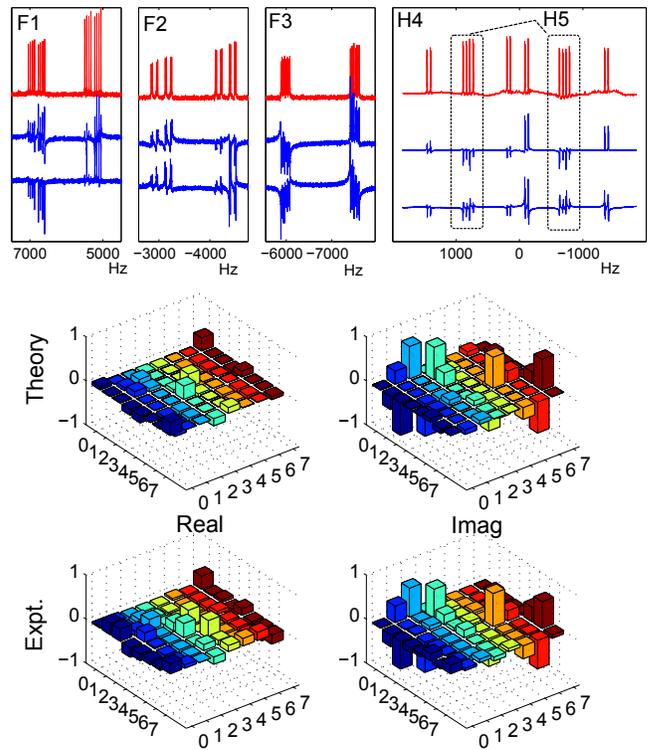} 
\caption{
AAQST results for the state $\rho_2$ described in the text.
The reference spectrum is in the top trace.  The
real (middle trace) and the imaginary spectra (bottom trace) 
are obtained in a single shot AAQST experiment.
The bar plots correspond to theoretically expected
states (top row) and those obtained from AAQST 
experiments (bottom row). Fidelity of the AAQST state is 0.95.
}
\label{btfbzres2} 
\end{figure}
\end{center}

The partially oriented spin-system yields all the 80 transitions
sufficiently resolved.  
Again we use just one experiment for the complete AAQST of the 3-qubit
input register.
We modelled the AAQST unitary in a similar way as before: 
$U_1 = \left(\frac{\pi}{2}\right)_x U_\mathrm{int}(\tau_2) 
\left(\frac{\pi}{2}\right)_x U_\mathrm{int}(\tau_1)$
where $U_\mathrm{int}(\tau) = \exp\left(-i{\cal H}\tau\right)$ is
the unitary operator for evolution under the internal Hamiltonian
${\cal H}$ (see eqn. (\ref{ham})) for a time $\tau$, and $\left(\frac{\pi}{2}\right)_x$ are 
global x-rotations.  The constraint matrix $M$ had 63 columns corresponding
to the unknowns and 160 rows corresponding to the real and imaginary
parts of 80 spectral lines.
After optimizing the durations by minimizing the condition number
using a genetic algorithm, we obtained $C(M) = 14.6$
for $\tau_1 = 431.2 ~\upmu$s and $\tau_2 = 511.5 ~\upmu$s.
Again we study AAQST on two states:
(i) Thermal equilibrium of the $^{19}$F spins: $\rho_1 = (\sigma_z^1+\sigma_z^2+\sigma_z^3)/2$, 
and 
(ii) a random density matrix $\rho_2$ obtained by applying unitary
$
U_0 = \left(\frac{\pi}{2}\right)_x^{F} \tau_0 (\pi)_x^{H} \tau_0 \left(\frac{\pi}{2}\right)_y^{F_1},
$ with $\tau_0 = 2.5$ ms, on thermal equilibrium state, i.e., $\rho_2 = U_0 \rho_1 U_0^\dagger$.
In both the cases, we initialize the ancilla i.e., the $^1$H qubits on to
a maximally mixed state by first applying a $(\pi/2)^{H}$ pulse followed
by a strong PFG in the $z$-direction.

The real and imaginary parts of the single shot AAQST spectra, 
along with the reference spectra, are shown in
Figs. \ref{btfbzres1} and \ref{btfbzres2} respectively.  
Again the line intensities
$\{R^{(1)}_{j\nu,j\nu'},S^{(1)}_{j\nu,j\nu'}\}$ are obtained by curve-fitting, and all the 63
unknowns of the 3-qubit deviation density matrix are obtained by solving the 
matrix eqn. (\ref{meq}). The reconstructed density matrices 
along with the theoretically expected states ($\rho_1$ and $\rho_2$)
are shown below the spectra 
in Figs. \ref{btfbzres1} and \ref{btfbzres2}. 
The fidelities of experimental states with
the theoretically expected states ($\rho_1$ and $\rho_2$)
are respectively 0.98 and 0.95.  The lower fidelity in the latter case
is mainly due the imperfections in the preparation of the target state
$\rho_2$.  The overall poorer performance in the liquid crystal system
is due to the lower fidelities of the QST pulses, spatial and temporal 
variations of solute order-parameter, and stronger decoherence rates
compared to the isotropic case.  In spite of these difficulties, 
the three-qubit density matrix with 63 unknowns could be 
estimated quantitatively through a single NMR experiment.

\section{Conclusions}
Quantum state tomography is an important part of experimental 
studies in quantum information processing.  The standard method 
involves a large number of independent measurements to reconstruct
a density matrix.  The ancilla-assisted quantum state tomography
introduced by Nieuwenhuizen and co-workers allows complete reconstruction
of complex density matrix with fewer experiments by letting the 
unknown state of the input register to interact with an ancilla
register initialized in a known state.  
Ancilla registers are essential in many of the quantum algorithms.
Usually, at the end of the quantum algorithms,
ancilla is brought to a state which is separable with the input 
register.  The same ancilla register which is used for computation
can be utilized for tomography after the computation. 
The ancilla register can be prepared into a maximally mixed state
by dephasing all the coherences and equalizing the populations.

We provided methods for
explicit construction of tomography matrices in large registers.  
We also discussed the optimization of tomography 
experiments based on minimization of the condition number of the 
constraint matrix.  Finally, we demonstrated the experimental 
ancilla-assisted quantum state tomography in two systems:  
(i) a system with two input qubits and one ancilla qubit in an isotropic medium
and (ii) a system with three input qubits and two ancilla qubits
in a partially oriented medium.  In both the cases, we 
successfully reconstructed the target density matrices 
with a single quadrature detection of transverse magnetization.  
The methods introduced in this work should be useful for extending 
the range of quantum state tomography to larger registers.

\section*{Acknowledgements}
The authors are grateful to Mr. Hemant Katiyar, Dr. Soumya S. Roy, and 
Prof. Anil Kumar for discussions.  This work was partly supported by 
DST project SR/S2/LOP-0017/2009.

\bibliographystyle{apsrev4-1}
\bibliography{bibfile1}

\end{document}